\documentclass[english,twoside,a4paper,10pt]{article}

\usepackage[latin1]{inputenc}
\usepackage[T1]{fontenc}
\usepackage{amsmath}
\usepackage{amsfonts}
\usepackage{graphicx}
\usepackage{subfigure}
\usepackage{a4wide}
\usepackage{amssymb}
\usepackage{fancyhdr}
\usepackage{mathrsfs}
\usepackage[toc,page]{appendix}

\begin{document}

\title{\bf Non-local $F(R)$-mimetic gravity}
\author{ 
Ratbay Myrzakulov\footnote{Email: rmyrzakulov@gmail.com},\,\,\,
Lorenzo Sebastiani\footnote{E-mail address: l.sebastiani@science.unitn.it
}\\
\\
\begin{small}
Department of General \& Theoretical Physics and Eurasian Center for
\end{small}\\
\begin{small} 
Theoretical Physics, Eurasian National University, Astana 010008, Kazakhstan
\end{small}\\
}

\date{}

\maketitle

%%%%%%%%%%%%%%%%%%%%%%%%%%%%%%%%%%%%%%%%%%%%%%%%%%%%%%%%%%%%%%%%%%%%%%%%%%%%%%%%%%%%%%%%%%%%%%%%%%%%%%%%%%%%%%%%%%%%%%%%%%%%%%%%%%%

%%%%%%%%%%%%%%%%%%%%%
%  Abstract
%%%%%%%%%%%%%%%%%%%%%
\begin{abstract}
In this paper, we study non-local $F(R)$-mimetic 
gravity. We implement mimetic gravity in the framework of non-local $F(R)$-theories of gravity. Given some specific class of models and using a potential on the mimetic field, we investigate some scenarios related to the early-time universe, namely the inflation and the cosmological bounce, which bring to Einstein's gravity with cold dark matter at the late-time. 
\end{abstract}
%%%%%%%%%%%%%%%%%%%%%

%----------------------------
%PACS
%----------------------------

%===========================================================================

\tableofcontents
%%%%%%%%%%%%%%%%%%%%%%%%%%%
%%%  Sec. I
%%%%%%%%%%%%%%%%%%%%%%%%%%%
\section{Introduction}

In this years, the interest in modified theories of gravity considerably
grew up due to the possibility to describe with them a huge variety of scenarios
in different cosmological contexts. In modified gravity new freedom degrees are introduced in the gravitational action of General Relativity, by replacing the Hilbert-Einstein term, namely the Ricci scalar, with a general function of the curvature invariants (see Refs.~\cite{R1, R2, CF, R3, R4, R5} for some reviews). 

In 2007 Deser \& Woodard~\cite{Desernonlocal}, inspired by quantum loop
corrections, proposed a new class of models based on non-local modifications to gravity, where the gravitational Lagrangian contains the inverse of some differential operator of the curvature invariants. The simplest choice,  also suggested by the dynamics of gravitational waves in the Friedmann-Robertson-Walker space-time,  is given by the inverse of the d'Alambertian operator of the Ricci scalar, and the theory is dubbed non-local $F(R)$-gravity. In Ref.~\cite{Odnonlocal} Nojiri and Odintsov investigated non-local $F(R)$-gravity for inflation and the dark energy epoch, showing that it is possible to achieve an unified description. The proprieties of the cosmological solutions in non-local gravity are analyzed in Refs.\cite{Liud, Koiv} and the initial conditions problem is investigated in Ref.~\cite{Calcagni}. 
Generalized non-local $F(R)$-gravity was proposed in Ref.~\cite{nl2}.
Other studies can be found in Refs.~\cite{nl1, nl3, nl4} and references therein.

Recently, a new approach to the dark matter issue has been proposed by
Chamseddin and Mukhanov: by isolating in a covariant way the conformal degree of freedom of 
the Einstein's theory 
via a singular disformal transformation of the metric, one gets new solutions
induced by a matter-like fluid~\cite{muk1, muk2}. If fact, the same result can be obtained by adding to the gravitational action a scalar field with norm equal to minus one~\cite{bekenstein, m1, m2, m3, m4}. Several works on mimetic gravity can be found in Refs.~\cite{m5}--\cite{m19}.

In our paper, we would like to recast mimetic gravity in non-local $F(R)$-modified gravity framework. Since this theories are mainly motivated by quantum effects at high energy, we will analyze some solutions related to early-time universe, namely
inflation and cosmological bounce
(see Refs.~\cite{Linde, revinflazione, Novello} for several reviews). We will also use a potential of the mimetic field, in order to modify its behaviour in the scenarios under investigation. For large values of the time, non-local gravity effects and mimetic potential disappear in order to recover Einstein's gravity with mimetic dark matter.

The paper is organized in the following way. In Section {\bf 2} we present the formalism of our model and the Equations of motion on Friedmann-Robertson-Walker space-time. Sections {\bf 3}--{\bf 4} are devoted to the study of inflation and comsological bounce in some relevant class of models of non-local $F(R)$-mimetic gravity. Conclusions are given in Section {\bf 5}.

%%% Unit %%%
We use units of $k_{\mathrm{B}} = c = \hbar = 1$ and we pose
$8\pi/M_{Pl}^2\equiv \kappa^2$, where $M_{Pl}$ is the Planck Mass.
%%%%%%%%%%%%
%%%%%

\section{The model}

We will consider the following gravitational action, where mimetic gravity~\cite{muk1,muk2} is implemented in non-local $F(R)$-gravity~\cite{Desernonlocal, Odnonlocal},
\begin{equation}
I=\int_\mathcal{M} d^4 x\sqrt{-g}
\left[
\frac{R}{2\kappa^2}\left[1+f(\Box^{-1} R)\right]+\lambda\left(
g^{\mu\nu}\partial_\mu\psi\partial_\nu\psi+1
\right)-V(\psi)
\right]\,.\label{action0}
\end{equation} 
Here, $\mathcal M$ is the space-time manifold, $g$ is the determinant of the metric tensor $g_{\mu\nu}$, $R$ is the Ricci scalar, $f(\Box^{-1} R)$ is a function of the inverse of the d'Alambertian operator $\Box\equiv\nabla_\rho\nabla^\rho$ of the curvature\footnote
{
In general
\begin{equation*}
\Box \equiv\frac{1}{\sqrt{-g}}\partial_\rho\left[\sqrt{-g}\partial^\rho \right]\,.
\end{equation*}
On flat Friedmann-Roberton-Walker space-time where $\sqrt{-g}=a(t)^3$, we get
\begin{equation*}
\Box^{-1} R\equiv -\int^t dt'\frac{1}{a(t')^3}\int^{t'} dt'' a(t'')^3 R(t'')\,,
\end{equation*}
and we see that the Lagrangian is non-local and in every position it is determined by the contour of the space-time.
},
 and $\psi$ is a scalar field subjected to the potential $V(\psi)$. The Lagrangian multiplier $\lambda$ introduces a constraint on the mimetic field $\psi$, such that, by taking the variation of the action respect to $\lambda$, one has,
\begin{equation}
g^{\mu\nu}\partial_\mu\psi\partial_\nu \psi=-1\,.\label{psiconst}
\end{equation}
To derive the field equations of the theory, it is convenient to introduce a scalar field $\phi$ with the Lagrangian multiplier $\xi$ to get a local scalar-tensor representation~\cite{Odnonlocal}, 
\begin{equation}
I=\int_\mathcal{M} d^4 x\sqrt{-g}
\left[
\frac{1}{2\kappa^2}\left[R(1+f(\phi))+\xi\left(\Box \phi-R\right)\right]+\lambda\left(
g^{\mu\nu}\partial_\mu\psi\partial_\nu\psi+1
\right)-V(\psi)
\right]\,.
\end{equation} 
The variation respect to $\xi$ leads to
\begin{equation}
\Box \phi=R\,,
\end{equation}
and one recovers (\ref{action0}). The field equations follow from the variation respect to the metric tensor $g_{\mu\nu}$ together with (\ref{psiconst}),
\begin{eqnarray}
&&\hspace{-3cm}R_{\mu\nu}(1+f(\phi)-\xi)-\frac{1}{2}g_{\mu\nu}\left[R(1+f(\phi)-\xi)-\partial_\rho\xi\partial^\rho\phi\right]=\frac{1}{2}(\partial_\mu\xi\partial_\nu\phi+\partial_\mu\phi\partial_\nu\xi)
\nonumber\\&&
-(g_{\mu\nu}\Box-\nabla_\mu\nabla_\nu)(f(\phi)-\xi)-2\kappa^2\lambda\partial_\mu\psi\partial_\nu\psi-\kappa^2 g_{\mu\nu} V(\psi)\,,\label{fieldeq}
\end{eqnarray}
$R_{\mu\nu}$ being the Ricci tensor. From the derative respect to $\phi$ one has
\begin{equation}
\Box\xi+f'(\phi) R\,,
\end{equation}
where $f'(\phi)\equiv d f(\phi)/d\phi$. The variation respect to $\psi$ leads to
\begin{equation}
-\frac{1}{\sqrt{-g}}\partial_\nu\left(\sqrt{-g}\lambda\partial^\nu\psi\right)=\frac{1}{2}\frac{d V(\psi)}{d\psi}\,,\label{eqpsi}
\end{equation}
but this equation is automatically satisfied by (\ref{fieldeq}) with (\ref{psiconst}). By looking to the field equations (\ref{fieldeq}), we see that the contribute from $\psi$ can be read as the one of a perfect fluid whose stress-energy tensor is given by:
\begin{eqnarray}
T_{\mu\nu}=(\rho+p)\partial_\mu\psi\partial_\nu\psi+p g_{\mu\nu}\,,\quad \rho=-2\lambda+V(\psi)\,,\quad p=-V(\psi)\,,
\end{eqnarray}
$\partial_\mu\psi$ being the four-velocity of the fluid with norm equal minus one (thanks to the constraint (\ref{psiconst})). Thus, in the limit $V(\psi)=0$, we obtain a dark-matter like fluid, while the potential modify its behaviour.

In this paper, we will look for flat Friedmann-Robertson-Walker (FRW) metric,
\begin{equation}
ds^2=-dt^2+a(t)^2 d{\bf x}^2\,,
\end{equation}
where $a\equiv a(t)$ is the scale factor of the universe, which brings to the identification of the mimetic field with the cosmological time (up to an arbitrary constant),
\begin{equation}
\psi=t\,.
\end{equation}
 The field equations read
\begin{equation}
3H^2(1+f(\phi)-\xi)=\frac{1}{2}\dot\xi\dot\phi-3H(f'(\phi)\dot\phi-\dot\xi)
-2\kappa^2\lambda+\kappa^2 V(t)\,,\label{F}
\end{equation}
\begin{equation}
-\left(2\dot H+3H^2\right)(1+f(\phi)-\xi)=\frac{1}{2}\dot\xi\dot\phi+\left(\frac{d^2}{d t^2}+2H\frac{d}{d t}\right)(f(\phi)-\xi)
-\kappa^2 V(t)\,.
\end{equation}
Here, $H=\dot a/a$ is the Hubble parameter and the dot is the time derivative. The equations for $\xi$ and $\phi$ are

\begin{equation}
\ddot\phi+3H\dot\phi=-12H^2-6\dot H\,.\label{phi}
\end{equation}
\begin{equation}
\ddot\xi+3H\dot\xi=\left(12H^2+6\dot H\right)f'(\phi)\,,\label{xi}
\end{equation}
Moreover, the equation for $\psi$ in (\ref{eqpsi}) is
\begin{equation}
\dot\lambda+3H\lambda=\frac{\dot V(t)}{2}\,,\label{psi}
\end{equation}
where we have taken into account that $\psi=t$. We note that, when $V(t)=0$, $\lambda\propto1/a^3$. 

Following Ref.~\cite{nl2},
the general feature of the model in FRW space-time may be investigated by 
recasting the equations of motion in an autonomous system and by
studying the stationary points (and their stability). One can introduce the dynamical variables,
\begin{equation}
N=\log a(t)\,,\quad x=\frac{\dot\phi}{H}\,,\quad y=\frac{\dot\xi}{H f(\phi)}\,,\quad
z=\frac{\kappa^2\lambda}{H^2 f}
\end{equation}
which obey to the system
\begin{eqnarray}
\frac{d x}{d N}&=&-3x-12-\frac{\dot H}{H^2}(x+6)\,,\nonumber\\
\frac{d y}{d N}&=&-3y+(12-y x)\left(\frac{f'(\phi)}{f(\phi)}\right)
-\frac{\dot H}{H^2}\left(y-6\frac{f'(\phi)}{\phi}\right)\,,\nonumber\\
\frac{d z}{d N}&=&-3z-2z\frac{\dot H}{H^2}-z x\left(\frac{f'(\phi)}{f(\phi)}\right)+\frac{\kappa^2\dot V(t)}{2H^3 f(\phi)}\,,\label{sp}
\end{eqnarray}
with the constraint
\begin{equation}
\frac{\xi-1}{f(\phi)}=1-\frac{x y}{6}+\frac{f'(\phi)}{f(\phi)}x-y+\frac{2z}{3}-\frac{\kappa^2V(t)}{3H^2f(\phi)}\,.
\end{equation}
In the limit $z=0$, the first two equations in (\ref{sp}) admit stationary points at ($x,y$)=($0,0$) and $(x,y)=(-2f(\phi)/f'(\phi), 6f'(\phi)/(2f(\phi)-3f'(\phi))$. The first one describes the Minkowski space-time, while the second (if $f(\phi)\neq 0$) can be used to generate new solutions (like the de Sitter solution associated to the dark energy).
This kind of analysis reveals the evolution of the system in the limit of non-local gravity where the new degree of freedom associated to the mimetic nature of the field is not considered ($\lambda=0$). 

To study cosmological solutions with $\lambda\neq 0$ and, more in general, $V(\phi)\neq 0$, we need to take into account also the last equation in (\ref{sp}), such that the analysis becomes much more involved. Thus, in our work we will look only for some specific solutions of the field equations.
In particular, we will see which role plays the mimetic field in some cosmological scenarios related with the early-time universe. In order to do it, we will consider some suitable form of $f(\phi)$ with a transition phase at $\phi=\phi_0$, such that $f(\phi)\simeq 0$ when $|\phi|\ll |\phi_0|$. The approach is purely phenomenological and it is understood that, in the limits $|\phi|\ll |\phi_0|$ and $t\rightarrow +\infty$, Einstein's gravity with mimetic cold dark matter (namely $V(t)\rightarrow 0$) must be recovered.

\section{Exponential form of $f(\phi)=f_0\left( \text{e}^{\phi/\phi_0}-1\right)$}

Let us start with the following form for $f(\phi)$,
\begin{equation}
f(\phi)=f_0\left( \text{e}^{\phi/\phi_0}-1\right)\,,\label{initial}
\end{equation}
where $f_0$ is an adimensional constant. The non-local gravity effects appear at $|\phi_0|<|\phi|$, while $f(\phi)$ goes to zero when $|\phi|\ll |\phi_0|$. 
If we consider a (positive) constant de Sitter solution $H=H_\text{dS}$, 
in the limit $\phi_0\ll |\phi|$,
equations (\ref{phi})--(\ref{xi}) can be solved as
\begin{equation}
\phi(t)\simeq -4 H_\text{dS} t+\phi_1\,,\quad
\xi\simeq \frac{3\text{e}^{\frac{\phi_1-4H_\text{dS}t}{\phi_0}}f_0\phi_0}{4-3\phi_0}
-\frac{c_1\text{e}^{-3 H_\text{dS}t}}{3H_0}+c_2\,,\label{res1}
\end{equation}
where $c_{1,2}$ and $\phi_1$ are constants. We may set
\begin{equation}
\phi_1=\phi_0+4H_\text{dS} t_0\,,\quad
\phi(t)=\phi_0+4H_\text{dS}(t_0-t)\,,\label{field1}
\end{equation}
such that, by choosing $0<\phi_0$, the field is positive and much larger than $\phi_0$ when $t\ll t_0$, while $\phi(t)=\phi_0$ when $t=t_0$. From Eq.~(\ref{F}) one obtains
\begin{equation}
-\frac{3H_\text{dS}^2\text{e}^{-\frac{4H_\text{dS}t}{\phi_0}}
\left(
-2\text{e}^{\phi_1/\phi_0}f_0(\phi_0-2)+\phi_0\text{e}^{\frac{4H_\text{dS}t}{\phi_0}}
(c_2-1)
\right)
}{\phi_0}=-2\kappa^2\lambda+\kappa^2 V(t)\,,\label{last}
\end{equation}
with $\lambda$ given by (\ref{psi}).
Thus, in the absence of mimetic contribute ($\lambda=V(t)=0$), the de Sitter solution is realized for $\phi_0=2$ and $c_2=1$ and we recover the result of  Ref.~\cite{Odnonlocal}, but in general the presence of the mimetic sector may be important. 
In the inflationary scenario with $|\dot H|\ll H^2$ and $H$ almost a constant, (\ref{res1}) are still valid. Now, if we set $\phi_0=2$, Eq.~(\ref{last}) reads
\begin{equation}
3H^2(1-c_2)=-2\kappa^2\lambda+\kappa^2 V(t)\,.\label{eqH1}
\end{equation}
To support the de Sitter expansion, we may use the following potential,
\begin{equation}
V(t)=V_0\text{e}^{\gamma(t_\text{i}-t)}\,,\label{pot1}
\end{equation}
where $V_0$ and $\gamma$ are positive constants, and $t_\text{i}$ is the initial time of the early-time acceleration, such that when $t_\text{i}\ll t$ the potential vanishes.
From Eq.~(\ref{psi}), we get
\begin{equation}
\lambda=V_0\gamma\frac{\text{e}^{\gamma(t_\text{i}-t)}}{2(\gamma-3H_\text{dS})}-\rho_0\text{e}^{-3H_{\text{dS}} t}\,,
\end{equation}
with $(0<)\rho_0$ representing the energy density of a dark matter-like fluid. Therefore, by avoiding the contribute of $\rho_0$ in (\ref{eqH1}), the Hubble parameter is given by
\begin{equation}
H=H_\text{dS}^2\text{e}^{\gamma(t_\text{i}-t)}\,,\quad H_\text{dS}=
\frac{\gamma-c_2\gamma\pm\sqrt{(c_2-1)((c_2-1)\gamma^2-12V_0\kappa^2)}}
{6-6c_2}
\,.
\end{equation}
When $\gamma(t_\text{i}-t)\ll 1$, we find the de Sitter solution of inflation $H\simeq H_\text{dS}$, but during the time the Hubble parameter slowly decreases as
\begin{equation}
-\frac{\dot H}{H^2}=\frac{\gamma}{2H}\,.
\end{equation}
Thus, we must require $\gamma\ll H_\text{dS}$ (slow-roll regime). Acceleration ends when $H=\gamma/2$ after the time
$(t-t_\text{i})=-\log[\gamma^2/(4H_\text{dS}^2)]/\gamma$, which is large as long as $\gamma^2/H_\text{dS}^2$ is small. At the end of inflation $V(t)\simeq 0$ and the field $\psi$ plays the role of dark matter. During the de Sitter expansion, the field $\phi$ also decreases as in (\ref{field1}), and the non-local effects disappear when $\phi\ll \phi_0$: thus, $t_0$ may be identified as the time at the end of inflation, namely 
$t_0=t_\text{i}-\log[\gamma^2/(4H_\text{dS}^2)]/\gamma$.

More interesting is the case of $c_2=1$ in (\ref{last}). For the (quasi) de Sitter solution we obtain
\begin{equation}
\frac{6 f_0 H^2}{\phi_0}\text{e}^{\frac{\phi_1-4 H_\text{dS}t}{\phi_0}}(\phi_0-2)=
-2\kappa^2\lambda+\kappa^2 V(t)\,,\label{lastlast}
\end{equation}
and non-local gravity plays a dynamic role. Now we can introduce the potential,
\begin{equation}
V(t)=V_0\text{e}^{\frac{\phi_1-4 H_\text{dS}t}{\phi_0}}\,,\label{pot2}
\end{equation}
which brings to the solution of (\ref{psi}),
\begin{equation}
\lambda=\frac{2V_0\text{e}^{\frac{\phi_1-4 H_\text{dS}t}{\phi_0}}}{4-3\phi_0}-\rho_0\text{e}^{-3H_{\text{dS}} t}\,.
\end{equation}
As above, we can avoid the contribute of $\rho_0$ and infer from (\ref{lastlast}),
\begin{equation}
H^2=\frac{V_0\kappa^2\phi_0^2}{16f_0-20f_0\phi_0+6f_0\phi_0^2}\,,
\end{equation}
namely the de Sitter space-time is an exact solution of the model. However, this result depends on the limit $\phi_0\ll \phi$ in (\ref{initial}). When $\phi$ approach to $\phi_1$, we have a phase transition and in the limit $\phi\ll \phi_0$ one recovers the gravitational action of General Relativity with mimetic dark matter. A general remark is in order: the dependence on the time of $V(t)$ in (\ref{pot1}\,,\ref{pot2}) is independent on the FRW solution due to the identification $\psi=t$. Thus, at the late time, the potential goes to zero and the mimetic field $\psi$ mimics cold dark matter only.\\
\\
An other interesting solution related with the early-time universe is the cosmological bounce, where the Hubble parameter assumes the form
\begin{equation}
H=H_0(t-t_1)\,,\label{Hbounce}
\end{equation} 
$H_0$ being a (positive) constant and $t_1$ the time of the bounce: for $t<t_1$ the Hubble parameter is negative and we have a contraction, while for $t_1<t$ the Hubble parameter is positive and we have an expansion. 
The bounce scenario has been proposed as an alternative description respect to the Big Bang theory:
instead from an initial singularity, the universe emerges from a bounce.

We can study the behaviour of our model near to the bounce, when $t$ is close to $t_1$ and $H^2\ll \dot H$. Again, we take the exponential form of $f(\phi)$ given in (\ref{initial}) with the asymptotic limit $f(\phi)\simeq f_0 \text{e}^{\phi/\phi_0}\,, |\phi_0|\ll |\phi|$. Equations (\ref{phi})--(\ref{xi}) lead to
\begin{equation}
\phi(t)\simeq -3H_0 t^2+\phi_1+\phi_2 t\,,\quad
\xi\simeq \frac{6f_0\phi_0 H_0\text{e}^{\frac{\phi_2 t}{\phi_0}}}{\phi_2^2}+c_1+c_2 t\,,
\end{equation}
where we set
\begin{equation}
\phi_1=3H_0 t_1^2\,,\quad \phi\simeq3H_0(t_1^2-t^2)+\phi_2 t\,,
\end{equation}
in the derivation of $\xi$.
For simplicity, we take $3H_0 t_1\ll \phi_2$ and observe the following limits
\begin{equation}
\phi\simeq \phi_2 t\,, \quad \phi_1\ll \phi_2 t
\end{equation}
near to the bounce.
In these expressions, $\phi_2$, $c_{1,2}$ are positive constants.
Equation (\ref{F}) reads
\begin{equation}
\frac{1}{2}(6H_0 t-\phi_2)\left(
c_2+\frac{6 f_0 H_0\text{e}^{\frac{\phi_2 t}{\phi_0}}}{\phi_2}
\right)\simeq
-2\kappa^2\lambda+\kappa^2 V(t)\,.\label{eqbounce}
\end{equation}
To support the bounce we need a potential of the type
(in this prescription the dimension of $V_0$ is $[V_0]=[1/\kappa^2]$),
\begin{equation}
V(t)=V_0 \left[\text{e}^{\frac{\phi_2 t}{\phi_0}}(-\phi_2^2+3H_0\phi_0(\phi_0-(t-t_1)\phi_2))\right]\,,\label{V3}
\end{equation}
such that the solution of (\ref{psi}) reads
\begin{equation}
\lambda=
-\frac{V_0\phi_2^2\text{e}^\frac{\phi_2 t}{\phi_0}}{2}
-\rho_0\text{e}^{-\frac{3H_0 t(t-2t_1)}{2}}
\,,
\end{equation}
where the second term with $0<\rho_0$ represents the dark matter-like contribute.
If we consider $\phi_0\ll \phi_2 t$  and $3H_0 t\ll \phi_2$, we have that Eq.~(\ref{eqbounce}) with $\rho_0=0$ is satisfied by making the choice
\begin{equation}
V_0=-\frac{f_0}{\kappa^2\phi_0^2}\,.
\end{equation}
Here, $f_0$ and $V_0$ can be either positive or negative. The non-local gravity effects disappear when $\phi\ll \phi_0$. For example, if $0<\phi_0\ll 1$, when $0\ll (t^2-t_1^2)$, we have $\phi\simeq \phi_1$ at the time $t\simeq\phi_2/(3H_0)$. To recover the dark matter at large time, we must modify the potential (\ref{V3}) by introducing a phase transition. For instance,
\begin{equation}
V(t)\rightarrow V(t)\text{e}^{-\alpha (t-t_1)^m}\,,\label{shift}
\end{equation}
with $\alpha$ a positive parameter and $m$ an integer number larger than one, brings to $V(t)\simeq 0$ when 
$0\ll \alpha (t-t_1)^m$.

\section{Power-law form of $f(\phi)=\gamma\left(\frac{\phi}{\phi_0}\right)$}

In this section, we will work with power-law forms of $f(\phi)$. Let us start with
\begin{equation}
f(\phi)=\gamma\left(\frac{\phi}{\phi_0}\right)^n\,,
\end{equation}
where $\gamma$ is an adimensional constant and $n$ is a positive natural number. Non-local gravity effects appear in the limit $|\phi_0|\ll |\phi|$. 
Now we consider the de Sitter solution with positive $H=H_\text{dS}$.
As we did before, we take the following solution of (\ref{phi}),
\begin{equation}
\phi=-4 H_\text{dS} t+\phi_1\,,\label{phi22}
\end{equation}
with $\phi_1$ constant, 
such that from Eq.~(\ref{xi}) we derive
\begin{equation}
\xi=-c_1 \frac{\text{e}^{-3H_\text{dS}t}}{3H_\text{dS}}+c_2
+\frac{\gamma}{\phi_0^n}\sum_{\tilde n=1}^{n} c_{\tilde n}(n) \left(H_\text{dS} t\right)^{\tilde n}\,,
\end{equation}
where $c_{1,2}$ are constants and the coefficients $c_{\tilde n}(n)$ depend on the power of $f(\phi)$ and may contain the constant $\phi_1$. In the simple case $n=1$ one obtains
\begin{equation}
\xi=-c_1 \frac{\text{e}^{-3H_\text{dS}t}}{3H_\text{dS}}+c_2+\frac{4H_\text{dS}\gamma t}{\phi_0}\,.\label{xi22}
\end{equation}
Thus, Eq.~(\ref{F}) reads
\begin{equation}
-\frac{H_\text{dS}^2\left(3(c_2-1)\phi_0+\gamma(16+24H_\text{dS} t-3\phi_1)\right)}{\phi_0}=
-2\kappa^2\lambda+\kappa^2 V(t)\,.\label{last22}
\end{equation}
The de Sitter solution can be realized by choosing a suitable form of mimetic potential. For example, 
if we set $c_2=1$ and we take a potential linear respect to the time as,
\begin{equation}
V=V_0 t\,,\label{Vt}
\end{equation}
$V_0$ being a constant, the solution of (\ref{psi}) is given by
\begin{equation}
\lambda=\frac{V_0}{6H_\text{dS}}-\rho_0\text{e}^{-3H_{\text{dS}} t}\,.\label{lambdaVt}
\end{equation}
If we pose $\rho_0=0$ and $V_0=-24\gamma H_0^3/(\phi_0\kappa^2)$, we see that, under the choice $\phi_1=8$, Eq.~(\ref{last22}) is satisfied. Starting from this result, we may try to study solutions for inflation when $\dot H\ll H^2$ and (\ref{phi22}, \ref{xi22}) are still valid for the case $n=1$. Suppose to have $\phi_1$ positive and much larger than $\phi_0$, in order to work in non-local gravity regime. We also assume that, at the beginning of the early-time acceleration, $H_\text{dS} t\ll \phi_1$ (if $1<\phi_0\ll 1$, we can estimate $t=\phi_1/(4H_\text{dS})$ as the time necessary to have $\phi\simeq\phi_0$).
By taking $c_2=1$, from Eq.~(\ref{last22}) with (\ref{Vt})--(\ref{lambdaVt}), by neglecting the contribute from $\rho_0$, we get 
\begin{equation}
H^2\simeq H_\text{dS}^2+\frac{3V_0\kappa^2\phi_0(\phi_1-8)t}{\gamma(16-3\phi_1)^2}\,,
\quad H_\text{dS}^2=\frac{V_0^{2/3}\kappa^{4/3}\phi_0^{2/3}}{(48\gamma-9\gamma\phi_1)^{2/3}}\,,
\end{equation}
where we must require $V_0\kappa^2\phi_0 (\phi_1-8)/\gamma<0$.
The model may describe a (quasi) de Sitter solution in slow roll-regime as long as
\begin{equation}
-\frac{\dot H}{H_\text{dS}^2}=|\frac{9(\phi_1-8)}{(6\phi_1-32)}|\ll 1\,.
\end{equation}
During the time, the Hubble parameter decreases and the end of the accelerated phase corresponds to $-\dot H/H^2\simeq 1$. Moreover, also the field $\phi$ decreases and $f(\phi)\rightarrow 0$ when $\phi\ll \phi_0$. On the other hand, in order to recover mimetic dark matter, the potential $V(t)$ must be modified by introducing a phase transition at some time $t_\text{f}$ at the end of inflation, like for example
\begin{equation}
V(t)=V_0 t (1-\text{e}^{-\alpha t/t_\text{f}})\,,
\end{equation}
with $\alpha$ positive parameter. In this case, when $t\ll t_\text{f}$, the potential goes to zero and the field $\psi$ mimics the dark matter.\\ 
\\
Let us have a look for the bounce solution (\ref{Hbounce}). When $t$ is close to $t_1$ and $H^2\ll \dot H$, equations (\ref{phi})--(\ref{xi}) for $n=1$ lead to
\begin{equation}
\phi(t)\simeq -3H_0 t^2+\phi_1+\phi_2 t\,,\quad
\xi\simeq\frac{3 H_0\gamma t^2}{\phi_0}+c_1+c_2 t\,,
\end{equation}
where $c_{1,2}$ and $\phi_{1,2}$ are positive constants.
Equation (\ref{F}) reads
\begin{equation}
\frac{1}{2}(6H_0 t-\phi_2)\left(
c_2+\frac{6\gamma  H_0 t}{\phi_0}
\right)\simeq
-2\kappa^2\lambda+\kappa^2 V(t)\,.\label{eqbounce2}
\end{equation}
Therefore, we can use a potential in the form
\begin{equation}
V(t)=\frac{3}{2}H_0 t(t-2t_1)V_0\,,
\end{equation}
with $V_0$ constant, to support the bounce. In this case, from Eq.~(\ref{psi}) we derive, 
\begin{equation}
\lambda=\frac{V_0}{2}
-\rho_0\text{e}^{-\frac{3H_0 t(t-2t_1)}{2}}\,,
\end{equation}
and Eq.~(\ref{eqbounce2}) holds true if
\begin{equation}
V_0=\frac{12H_0\gamma}{\phi_0\kappa^2}\,,\quad c_2=\frac{-12H_0\gamma t_1+\gamma\phi_2}{\phi_0}\,,
\end{equation}
and
\begin{equation}
H_0=\frac{\phi_2^2}{12(2+t_1\phi_2)}\,.
\end{equation}
Like in the preceding examples, for large values of time, if $t$ grows up, the field $\phi$ decreases and non-local gravity effects disappear when $\phi\ll\phi_0$. On the other hand, to recover mimetic dark matter, the potential must be redefined as in (\ref{shift}), such that $V(t)\simeq 0$ when $(t-t_1)$ is large.\\
\\
Some final remarks are in order about the possibility to use mimetic non-local gravity to unify the early-time acceleration with the current dark energy epoch. Following the proposal of Ref.~\cite{Odnonlocal}, one may extend the model to $F(R)$-gravity introducing
a new freedom degree inside the theory to reproduce the dark energy. On the other hand, in our case, we can play with the potential $V(\phi)$, such that at the late time the mimetic field could support the acceleration. For example, potentials in the form $V(\phi)\sim 1/\phi^2$ can mimic quintessence~\cite{muk2}.

\section{Conclusions}

In this paper, we have implemented mimetic gravity in non-local $F(R)$-gravity framework. 
We worked in a local-tensor framework by using an auxiliary field with a Lagrange multiplier.
We have shown with some examples how this kind of models can be used to reproduce some scenarios related to early-time universe, namely inflation and cosmological bounce, compatibly with Einstein's gravity with mimetic dark matter at the late-time limit. The introduction of a potential of the mimetic field gives the possibility to realize the desired solutions. We remember that, thanks to the identification of the mimetic field with the cosmological time on Friedmann-Robertson-Walker space-time, the behaviour of the potential is independent on the specific cosmological solutions.

Other works on modified theories of gravity for inflation and cosmological bounce can be found in Refs.~\cite{DeFelice}--\cite{sei}. For alternative approach to dark matter see also Ref.~\cite{mioDM}.

%%%%%%%%%%%%%%%%%%%%BIBLIO

\end{document}